\documentclass[aps, prl, twocolumn, showkeys]{revtex4-1}

\usepackage{chemformula} 
\usepackage{siunitx} 
\usepackage[inline,shortlabels]{enumitem} 
\usepackage{hyperref} 
\usepackage{graphicx} 
\usepackage{tabularx} 

\usepackage{natbib}

\begin{document}

    \title{Dominant role of orbital splitting in determining cathode potential in $O3$  \ch{NaTMO2} compounds}
    \author{M. H. N. Assadi}
    \email{h.assadi.2008@ieee.org}
    \affiliation{Center for Computational Sciences, University of Tsukuba, Tennodai 1-1-1, Tsukuba, Ibaraki 305-8577, Japan.}
    \author{Y. Shigeta}
    \affiliation{Center for Computational Sciences, University of Tsukuba, Tennodai 1-1-1, Tsukuba, Ibaraki 305-8577, Japan.}
    \date{2018}

    \begin{abstract}
        Designing high potential cathodes for Na-ion batteries, which are comparable in performance to Li-ion cathodes, remains a challenging task. Through comprehensive density functional calculations, we disentangle the relationship between the cathode potential and the ionicity of  \ch{TM}--\ch{O} bonds in $O3$ \ch{NaTMO2} compounds in which \ch{TM} ions is a fourth- or fifth-row transition metal. We demonstrate that the magnetic exchange interaction and the local distortions in the coordination environment of \ch{TM} ions play more significant roles in determining the cathode potential of the \ch{TM^{3+} -> TM^{4+} + e^-} reaction than the ionicity of the \ch{TM}--\ch{O} bonds in these compounds. These results indicate that designing cathode materials solely based on empirical electronegativity values to achieve high potential may not be a feasible strategy without taking into account a detailed structural assessment.
    \end{abstract}
   \maketitle

    \section{Introduction}

\begin{figure}
            \centering
            \includegraphics[width=0.9\columnwidth]{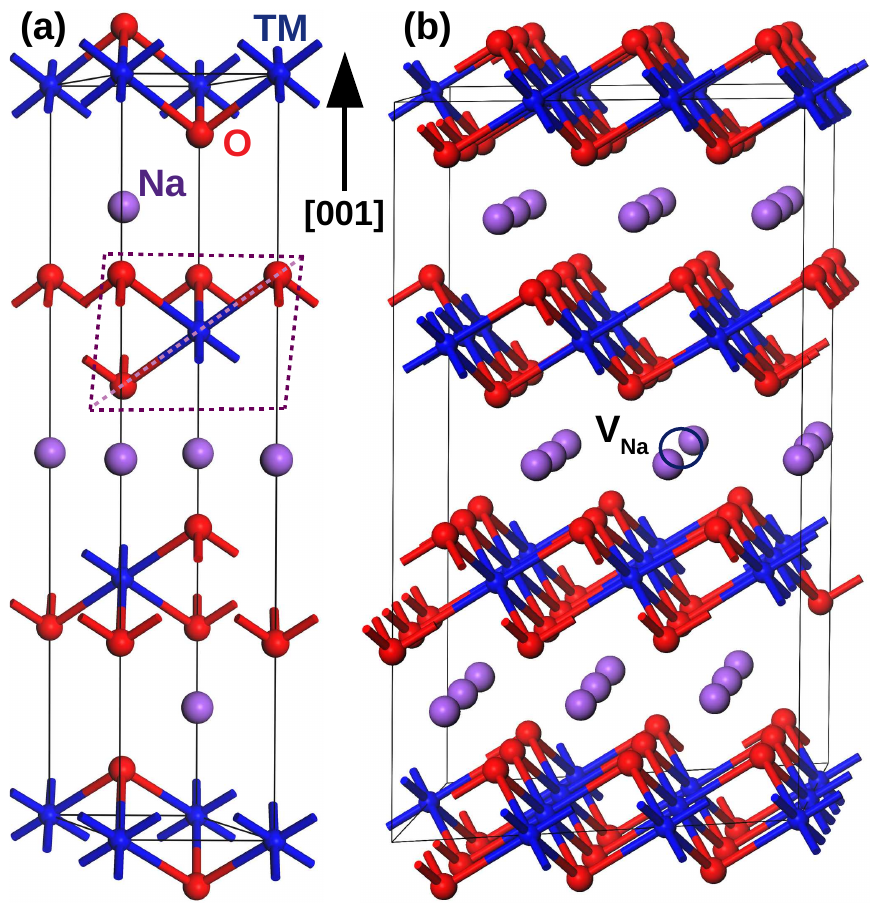}
            \caption{\label{fig:1}The hexagonal representation of the $O3$ \ch{NaTMO2} structure with $R\bar3 m$ symmetry (space group 166) is presented in (a). \ch{Na}, \ch{TM} and \ch{O} atoms occupy $3b$, $3a$ and $6c$ Wyckoff positions, respectively. The left panel (b) shows the $3a\times3a\times1c$ supercell used for calculating the sodium vacancy formation energy and cathode electrochemical potential.}
        \end{figure}

Layered transition metal (\ch{TM}) oxides constitute an important materials category for cathode applications in rechargeable batteries. The fundamental operating principle for such cathodes is based on the availability of multiple oxidation states for the \ch{TM} ion, which allows the removal or insertion of alkali metal atom from or into the framework of \ch{TM} oxide while maintaining the materials’ integrity. Although \ch{Li} atom is the most popular alkali metal atom in battery applications, the more affordable \ch{Na} atom has been considered a suitable alternative for Li atom. \ch{Na} atom, however, faces some challenges that need to be addressed before wide-scale industrial adaptation for portable applications becomes feasible \cite{Xiang2015}. For instance, \ch{Na} atom has a larger ionic radius and smaller ionisation potential compared to those of \ch{Li} atom resulting in lower performance \cite{Palomares2012, Liu2016}. Consequently, to design competitive \ch{Na} ion cathodes, all factors affecting the performance should be well understood and critically fine-tuned. One of such factors is the choice of \ch{TM} that would possibly maximise the electrochemical potential of the cathode.
It is generally speculated that the cathode potential is strongly correlated with the ionicity of the \ch{TM}--\ch{O} bond \cite{Liu2016, Dompablo2006, Barpanda2014}. The more ionic the \ch{TM}--\ch{O} bond is, the stronger valence electrons are attracted by the \ch{TM} nuclei. A stronger attraction, in turn, corresponds to higher energy for electron transfer, resulting in higher electrochemical potential upon the removal or the insertion of the alkali atoms \cite{Melot2013}. In this communication, we examine this conjecture in detail to see if this correlation holds in layered sodium \ch{TM} oxides of the $O3$ structure. Layered $O3$ structure, as shown in FIG. \ref{fig:1}(a) is made of three alternating \ch{TMO2} and Na layers arranged in hexagonal symmetry in which both \ch{TM} and Na ions are octahedrally coordinated by oxygen. We chose this class of materials for our investigation because they are among the most widely studied and commercialised \ch{Na} ion cathode materials. We also considered all 3d and 4d \ch{TM} ions for which $+3$ oxidation state is available.

    \section{Methods}

Spin-polarized density functional theory (DFT) calculations were carried out using augmented plane-wave method as implemented in VASP \cite{Kresse1996a, Kresse1996b}. The energy cut-off was set to 550 \si{eV}, while a $k$-point mesh was produced by Monkhorst Pack scheme with a spacing of $\sim 0.02$ \si{\per\angstrom} for Brillouin zone sampling. Hubbard term \cite{Dudarev1998} ($U_{eff}$) was added to the 3d and 4d electrons to improve the accuracy of the electronic description through GGA $+U$ approach. The $U_{eff}$ values were 5 \si{eV} for 3d electrons and 2\si{eV} for 4d electrons, respectively. Lattice parameters and magnetic ordering of the \ch{TM} ions of all compounds was fixed to the previously established ground state values \cite{Assadi2018a}. A $3a\times 3a\times 1c$ supercell was used which is presented in FIG. \ref{fig:1}(b) to calculate the formation energy ($E_f$) of the sodium vacancy (\ch{V_{Na}}) and cathode potential. Using a relatively large supercell minimises the artificial vacancy-vacancy interaction that is caused by periodic boundary conditions \cite{Puska1998}.

    \section{Reslts and discussions}
    
\begin{figure*}
                \centering
                \includegraphics[width=1.9\columnwidth]{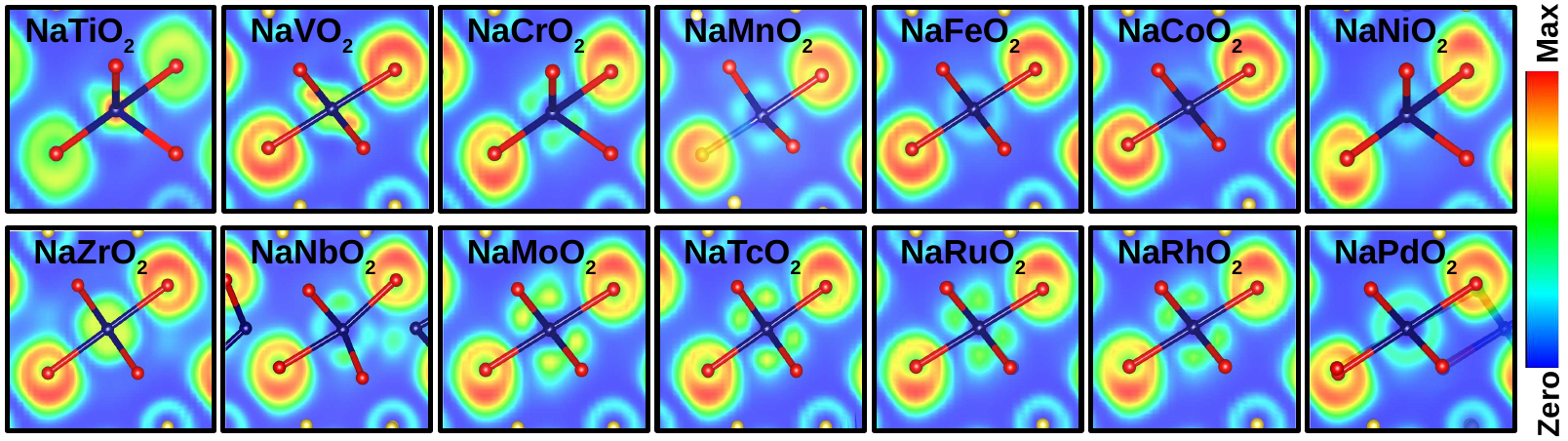}
                \caption{\label{fig:2}The electronic localisation function plotted for planes containing \ch{O}--\ch{TM}--\ch{O} bonds stretched from top right to left bottom of each panel. This plane is marked with dashed lines in FIG. \ref{fig:1}(a). Red corresponds to the maximum ELF values presented in Table \ref{table:1} while blue corresponds to zero ELF.}
            \end{figure*}

    We used a single \ch{V_{Na}} to simulate the early desodiation process. The cell voltage, $V$, is usually calculated using the following formula \cite{Islam2014}:
   \small       
            \begin{equation}\label{eq:1}
                V = \frac{-\left\{E_t(\ch{Na_yTMO2}) - E_t(\ch{Na_xTMO2}) - (y-x) E_t(\ch{Na})\right\}}{(y-x)e}.
            \end{equation}
   \normalsize        
Here, $E_t$ is the total energy of a given compound obtained by DFT calculations with the Hubbard $U_{eff}$ correction. \ch{Na_yTMO2} and \ch{Na_xTMO2} are the sodiated initial and desodiated final compounds, respectively. On the other hand, $E_f$ of creating one sodium vacancy in the supercell is given by the following standard equation \cite{Walle2004}:
   \small   
            \begin{equation}\label{eq:2}
                E_f = E_t (\ch{Na_nTM_nO_{2n}}) - E_t (\ch{Na_{n-1}TM_nO_{2n}}) - E_t (\ch{Na}).
            \end{equation}
   \normalsize        
By comparing Eq. \ref{eq:1} with Eq. \ref{eq:2}, we find that early in the desodiation process simulated by removing one Na atom from the supercell presented in FIG. \ref{fig:1}, the following relationship holds: $V =-(E_f (\ch{V_{Na}}))⁄e$. This potential is presented in Table \ref{table:1}. Along with the calculated values, the available experimental potentials for some of the compounds are also presented in Table \ref{table:1}. The differences between the experimental and the calculated values are within the range of $\sim 0.5$ \si{V} which is satisfactorily accurate. By only considering the early desodiation process, that is removing only one Na atom from the supercell, we could avoid taking account of the complex and successive phase transitions during sodiation or desodiation processes. These phase transitions usually differ from compound to compound which can hinder a veracious theoretical assessment of the chemical trend \cite{Xiang2015}. Furthermore, the removal of one Na atom guarantees that in all supercells one \ch{TM^3+} ion converts to \ch{ TM^4+} ion. This approach, therefore, offers us a straightforward insight into how the ionicity of the compound and the choice of \ch{TM} ion affect the cathode potential while keeping all other possible variables constant.
The ionicity of the \ch{TM}--\ch{O} bonds was determined by examining the electronic localisation function (ELF) \cite{Becke1990}. ELF is defined as the probability of finding a second like-spin electron near a given point. In the case of \ch{NaTMO2} compounds, large ELF values ($1 \textendash \sim0.6$) peaking around the ionic centres is a characteristic of ionic bonding, while large EFL peaking in the area connecting two ions implies covalent bonding. The calculated ELF is plotted in FIG. \ref{fig:2} for all compounds, while the maximum ELF values ($ELF_{Max}$) for the given compounds are presented in Table \ref{table:1}. The ELF was plotted along a plane containing an \ch{O}--\ch{TM}--\ch{O} bond as marked by dashed lines in FIG. \ref{fig:1}(a). No compound was found to have $ELF_{Max}$ smaller than $\sim 0.62$. Furthermore, ELF was larger around ionic centres, especially \ch{O}, and decreased in the bond regions. These factors indicate that the \ch{TM}--\ch{O} in all studied compounds, with varying strength, were ionic.

$ELF_{Max}$ values, presented in Table \ref{table:1}, show that there is no clear trend relating the ionicity of the \ch{TM}--\ch{O} bonds in 3d \ch{TM} containing \ch{NaTMO2} compounds to the \ch{TM} ions in the compound. For instance, \ch{NaTiO2}, \ch{NaMnO2} and \ch{NaNiO2} are more ionic than their neighbouring compounds. However, the ionicity of 4d containing \ch{NaTMO2} compounds monotonically increases as the atomic number of the \ch{TM} ions increases. Additionally, there is no clear correlation between the cathode potential and the ionicity of the \ch{TM}--\ch{O} bonds either. The most obvious case is \ch{NaTiO2} which has a negative potential value indicating the instability of \ch{Ti^3+} ($t_{2g}^1 e_g^0$) and its preference for adopting \ch{Ti^4+} ($t_{2g}^0 e_g^0$), though it has the most ionic \ch{TM}--\ch{O} bond among all compounds. Consequently, \ch{NaTiO2} is indeed suitable for anode rather than for cathode application \cite{Wu2015}. Furthermore, \ch{NaMnO2} has a \ch{TM}--\ch{O} bond more ionic than both neighbouring \ch{NaCrO2} and \ch{NaFeO2}, but its potential is nonetheless smaller than the potentials of both those compounds. In this case, the creation of the \ch{V_{Na}} transforms \ch{Mn^3+} ($t_{2g}^3 e_g^1$) to \ch{Mn^4+} ($t_{2g}^3 e_g^0$). Similarly, \ch{NaNiO2} is also more ionic than the neighbouring\ch{ NaCoO2}, although its potential is smaller than that of \ch{NaCoO2}. In this case, upon the removal of one Na atom, a \ch{Ni^3+} ($t_{2g}^6 e_g^1$) is transformed into \ch{Ni^4+} ($t_{2g}^6 e_g^0$). For 4d \ch{TM} containing compounds, the cathode potential did not follow the monotonic trend that governed the ionicity of the \ch{TM}--\ch{O} bonds with respect to the atomic number of the \ch{TM} ions; \ch{NaTcO2} and \ch{NaPdO2} are both more ionic than the preceding \ch{NaMoO2}, and \ch{NaRhO2} respectively, but still have smaller cathode potential. The desodiation process transforms \ch{Tc^3+} ($t_{2g}^3 e_g^1$) to \ch{Tc^4+} ($t_{2g}^3 e_g^0$) in the first compound and \ch{Pd^3+} ($t_{2g}^6 e_g^1$) to \ch{ Pd^4+} ($t_{2g}^6 e_g^0$) in the latter compound.

                \begin{figure*}
                \centering
                \includegraphics[width=1.9\columnwidth]{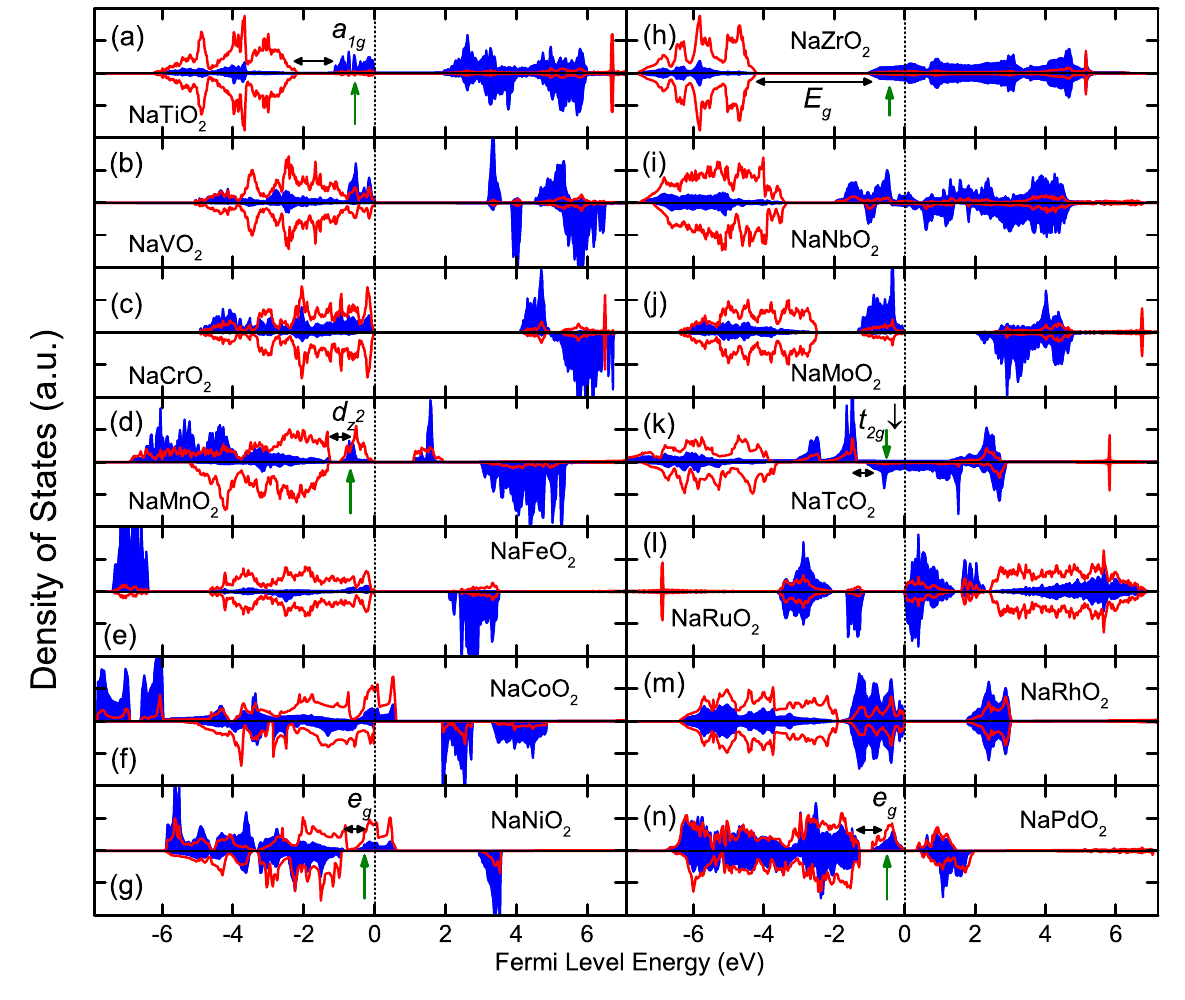}
                \caption{\label{fig:3}The site projected density of states of a single \ch{TMO2} chain in $O3$ \ch{NaTMO2} compounds. The blue shaded areas correspond to the d states of a single \ch{TM^3+} while the red lines correspond to the p states of the coordinating O ions.}
            \end{figure*}
            
Based on the results obtained, we can so far conclude that when the desodiation brings a \ch{TM} ion to an empty, half-filled or filled $t_{2g}$ configurations ($t_{2g}^0 e_g^0$, $t_{2g}^3 e_g^0$, or $t_{2g}^6 e_g^0$) the cathode potential experiences a reduction compared to that of neighbouring \ch{TM} ions in the same row. Furthermore, it also has become evident that the \ch{TM}--\ch{O} ionicity is not a good predictor of the cathode potential. But why is it so? The redox in \ch{NaTMO2} compounds is accomplished by the \ch{TM} ions donating an electron upon the extraction of a Na atom. As a result, it is only the last valence electron with the highest energy level that determines the cathode potential. If such electron has a significantly higher energy level than the rest of the valence electrons, then the cathode potential will be relatively smaller even if the compound is overall highly ionic. This is because the ionicity of the \ch{TM}--\ch{O} bond is based on the overall electron affinity of the entire valence electrons to their respective ionic centres \cite{Bergmann1987, Politzer2002}.

Now. Let’s examine what characteristics of the highest energy electron results in smaller potential. Examining the site projected density of states as presented in FIG. \ref{fig:3} reveals that when the last occupied \ch{TM^3+} electron comes from a singly occupied orbital which is detached from the main valence band, the cathode potential is relatively smaller regardless of the ionicity of the \ch{TM}--\ch{O} bond. These detached orbitals are marked with green arrows for \ch{NaTiO2}, \ch{NaMnO2}, \ch{NaNiO2}, \ch{NaTcO2} and \ch{NaPdO2} in FIG. \ref{fig:3}. The detached orbitals can be created either by a magnetic exchange splitting in elements for which one spin channel is singly occupied or by a lattice distortion which splits the occupied states or simply in \ch{TM} ions with single electron such as \ch{Ti^3+} ion. The magnetic exchange splitting is a natural result of the spin-dependent Hamiltonian of the system. If the exchange splitting is larger than the spin channels’ bandwidth, and the spin minority channel is singly occupied, then the last occupied orbital becomes detached from the main valence band. This is the case in \ch{NaTcO2}. In \ch{NaMnO2}, a strong Jahn-Teller distortion \cite{Assadi2018a, Jia2011} favours high-spin configuration. However, the crystal field splitting of the spin-up $e_g$ states is large enough that prevents the hybridisation of $e_{z^2}$ orbital with the rest of the valence band. In \ch{NaNiO2} and \ch{NaPdO2}, the octahedral crystal field separates the fully occupied $t_{2g}$ orbitals from the singly occupied $e_g$ orbital.

In the compounds mentioned above, the detachment of the singly occupied orbitals is large enough that it creates a pseudo-gap within the valence band. As a result, the hybridisation between the highest energy \ch{TM} electron and \ch{O} 2p states which tend to gravitate towards the bottom of the valence band is significantly reduced. This lack of hybridisation further lowers the energy required for the redox reaction. This pseudo-gap is $1.112$ \si{eV} in \ch{NaTiO2}, $0.311$ \si{eV} in \ch{NaMnO2}, $0.422$ \si{eV} in \ch{NaNiO2}, $0.333$ \si{eV} in \ch{NaTcO2} and $0.374$ \si{eV} in \ch{NaPdO2}. In the case of \ch{NaZrO2} in which a \ch{Zr^3+} ($t_{2g}^1 e_g^0$) transforms to a \ch{Zr^4+} ($t_{2g}^0 e_g^0$), the donated electron leaves the conduction band (also marked with a green arrow) instead of the valence band. In this case, the donated electron comes from orbitals higher in energy by the fundamental bandgap, i.e. $3.222$ \si{eV}, than the bottom of the valence band.

As a final note, we would like to mention that according to Table \ref{table:1}, the cathode potential is generally smaller for the 4d \ch{TM} containing compounds than those for the 3d \ch{TM} containing compounds. The smaller potential for 4d \ch{TM} ions is typically expected as with increasing period; the more modest localisation effects lead to the weaker attraction between the electrons and the \ch{TM} nuclei. Therefore, the use of 3d \ch{TM} ions is preferred over the use of 4d \ch{TM} ions for achieving high potentials.

           \begin{table}
           \caption{Maximum values for electronic localisation function ($ELF_{Max}$), the calculated cathode potential. When experimental data was available, the experimental cathode potential is also cited. The experimental potential corresponds to compounds that are nearly fully sodiated.}
 \label{table:1} 
        \begin{tabularx}{220pt}{l c X X c}
\hline
\hline
 System & $ELF_{Max}$ & Calculated Potential (\si{V}) & Experimental Potential (\si{V}) & Ref.  \\
 \hline
\ch{NaTiO2} & $0.9360$ & $-0.438$ &  & \\
\ch{NaVO2} & $0.6446$ & $\phantom{-}2.127$ & $\sim1.8$ & \cite{Didier2011}\\
\ch{NaCrO2} & $0.6843$ & $\phantom{-}3.610$ & $\sim 3$ & \cite{Yu2015, Xia2011}\\
\ch{NaMnO2} & $0.6941$ & $\phantom{-}2.146$ & $\sim 2.5$ & \cite{Ma2011}\\
\ch{NaFeO2} & $0.6475$ & $\phantom{-}3.004$ & $\sim 3.4$ & \cite{Yabuuchi2012, Zhao2013}\\
\ch{NaCoO2} & $0.6426$ & $\phantom{-}4.129$ & $\sim 4$ & \cite{Shibata2015}\\
\ch{NaNiO2} & $0.7120$ & $\phantom{-}1.987$ & $\sim 2$ & \cite{Vassilaras2013}\\
\ch{NaZrO2} & $0.6524$ & $-1.021$ &  & \\
\ch{NaNbO2} & $0.6881$ & $\phantom{-}1.072$ &  & \\
\ch{NaMoO2} & $0.6907$ & $\phantom{-}1.820$ & $\sim 1.1$ & \cite{Vitoux2017}\\
\ch{NaTcO2} & $0.6916$ & $\phantom{-}0.617$ &  & \\
\ch{NaRuO2} & $0.7023$ & $\phantom{-}2.067$ &  & \\
\ch{NaRhO2} & $0.7034$ & $\phantom{-}2.409$ &  & \\
\ch{NaPdO2} & $0.7123$ & $\phantom{-}2.306$ &  & \\
\hline
\hline
        \end{tabularx}
          \end{table}

    \section{Conclusions}
       In conclusion, we demonstrated when the extraction of a Na atom brings a \ch{TM} ion to one of the $t_{2g}^0 e_g^0$, $t_{2g}^3 e_g^0$ or $t_{2g}^6 e_g^0$ configurations under octahedral coordination cathode potential is smaller than the case in which the final \ch{TM} electronic configuration is otherwise. That is because of the last electron in \ch{TM^3+} ion is detached from and less hybridised with the rest of the valence band orbitals. Expanding this concept to tetrahedral symmetry, we anticipate similarly that if compounds in which \ch{TM} electronic configuration after desodiation was $e^0 t_2^0$, $e^2 t_2^0$, or $e^4 t_2^0$, the cathode potential would be potentially smaller. This electronic consideration has a stronger effect on determining the cathode potential than the ionicity of the \ch{TM}--\ch{O} bonds. At last, it is worth noting that our conclusions only hold if the redox reaction is fully compensated by the \ch{TM} ion electrons and may not be readily generalised to compounds in which \ch{O} ions also contribute to charge compensation during the redox reaction \cite{Assadi2018b}.
        
  \section{acknowledgments}
       This work was supported in part by MEXT as a social and scientific priority issue: Creation of new functional devices and high-performance materials to support next-generation industries to be tackled by using post-K Computer. Computational resources were provided by Kyushu University’s high-performance computing centre and supercomputers at the Institute for Solid State Physics at the University of Tokyo and the Centre for Computational Sciences at the University of Tsukuba.
        
   \bibliography{Ref}{}

\begin{thebibliography}{28}%
\makeatletter
\providecommand \@ifxundefined [1]{%
 \@ifx{#1\undefined}
}%
\providecommand \@ifnum [1]{%
 \ifnum #1\expandafter \@firstoftwo
 \else \expandafter \@secondoftwo
 \fi
}%
\providecommand \@ifx [1]{%
 \ifx #1\expandafter \@firstoftwo
 \else \expandafter \@secondoftwo
 \fi
}%
\providecommand \natexlab [1]{#1}%
\providecommand \enquote  [1]{``#1''}%
\providecommand \bibnamefont  [1]{#1}%
\providecommand \bibfnamefont [1]{#1}%
\providecommand \citenamefont [1]{#1}%
\providecommand \href@noop [0]{\@secondoftwo}%
\providecommand \href [0]{\begingroup \@sanitize@url \@href}%
\providecommand \@href[1]{\@@startlink{#1}\@@href}%
\providecommand \@@href[1]{\endgroup#1\@@endlink}%
\providecommand \@sanitize@url [0]{\catcode `\\12\catcode `\$12\catcode
  `\&12\catcode `\#12\catcode `\^12\catcode `\_12\catcode `\%12\relax}%
\providecommand \@@startlink[1]{}%
\providecommand \@@endlink[0]{}%
\providecommand \url  [0]{\begingroup\@sanitize@url \@url }%
\providecommand \@url [1]{\endgroup\@href {#1}{\urlprefix }}%
\providecommand \urlprefix  [0]{URL }%
\providecommand \Eprint [0]{\href }%
\providecommand \doibase [0]{http://dx.doi.org/}%
\providecommand \selectlanguage [0]{\@gobble}%
\providecommand \bibinfo  [0]{\@secondoftwo}%
\providecommand \bibfield  [0]{\@secondoftwo}%
\providecommand \translation [1]{[#1]}%
\providecommand \BibitemOpen [0]{}%
\providecommand \bibitemStop [0]{}%
\providecommand \bibitemNoStop [0]{.\EOS\space}%
\providecommand \EOS [0]{\spacefactor3000\relax}%
\providecommand \BibitemShut  [1]{\csname bibitem#1\endcsname}%
\let\auto@bib@innerbib\@empty
\bibitem [{\citenamefont {Xiang}\ \emph {et~al.}(2015)\citenamefont {Xiang},
  \citenamefont {Zhang},\ and\ \citenamefont {Chen}}]{Xiang2015}%
  \BibitemOpen
  \bibfield  {author} {\bibinfo {author} {\bibfnamefont {X.}~\bibnamefont
  {Xiang}}, \bibinfo {author} {\bibfnamefont {K.}~\bibnamefont {Zhang}}, \ and\
  \bibinfo {author} {\bibfnamefont {J.}~\bibnamefont {Chen}},\ }\href {\doibase
  10.1002/adma.201501527} {\bibfield  {journal} {\bibinfo  {journal} {Adv.
  Mater.}\ }\textbf {\bibinfo {volume} {27}},\ \bibinfo {pages} {5343}
  (\bibinfo {year} {2015})}\BibitemShut {NoStop}%
\bibitem [{\citenamefont {Palomares}\ \emph {et~al.}(2012)\citenamefont
  {Palomares}, \citenamefont {Serras}, \citenamefont {Villaluenga},
  \citenamefont {Hueso}, \citenamefont {Carretero-Gonzalez},\ and\
  \citenamefont {Rojo}}]{Palomares2012}%
  \BibitemOpen
  \bibfield  {author} {\bibinfo {author} {\bibfnamefont {V.}~\bibnamefont
  {Palomares}}, \bibinfo {author} {\bibfnamefont {P.}~\bibnamefont {Serras}},
  \bibinfo {author} {\bibfnamefont {I.}~\bibnamefont {Villaluenga}}, \bibinfo
  {author} {\bibfnamefont {K.~B.}\ \bibnamefont {Hueso}}, \bibinfo {author}
  {\bibfnamefont {J.}~\bibnamefont {Carretero-Gonzalez}}, \ and\ \bibinfo
  {author} {\bibfnamefont {T.}~\bibnamefont {Rojo}},\ }\href {\doibase
  10.1039/C2EE02781J} {\bibfield  {journal} {\bibinfo  {journal} {Energy
  Environ. Sci.}\ }\textbf {\bibinfo {volume} {5}},\ \bibinfo {pages} {5884}
  (\bibinfo {year} {2012})}\BibitemShut {NoStop}%
\bibitem [{\citenamefont {Liu}\ \emph {et~al.}(2016)\citenamefont {Liu},
  \citenamefont {Neale},\ and\ \citenamefont {Cao}}]{Liu2016}%
  \BibitemOpen
  \bibfield  {author} {\bibinfo {author} {\bibfnamefont {C.}~\bibnamefont
  {Liu}}, \bibinfo {author} {\bibfnamefont {Z.~G.}\ \bibnamefont {Neale}}, \
  and\ \bibinfo {author} {\bibfnamefont {G.}~\bibnamefont {Cao}},\ }\href
  {\doibase https://doi.org/10.1016/j.mattod.2015.10.009} {\bibfield  {journal}
  {\bibinfo  {journal} {Materials Today}\ }\textbf {\bibinfo {volume} {19}},\
  \bibinfo {pages} {109} (\bibinfo {year} {2016})}\BibitemShut {NoStop}%
\bibitem [{\citenamefont {Arroyo-de Dompablo}\ \emph
  {et~al.}(2006)\citenamefont {Arroyo-de Dompablo}, \citenamefont {Armand},
  \citenamefont {Tarascon},\ and\ \citenamefont {Amador}}]{Dompablo2006}%
  \BibitemOpen
  \bibfield  {author} {\bibinfo {author} {\bibfnamefont {M.~E.}\ \bibnamefont
  {Arroyo-de Dompablo}}, \bibinfo {author} {\bibfnamefont {M.}~\bibnamefont
  {Armand}}, \bibinfo {author} {\bibfnamefont {J.~M.}\ \bibnamefont
  {Tarascon}}, \ and\ \bibinfo {author} {\bibfnamefont {U.}~\bibnamefont
  {Amador}},\ }\href {\doibase https://doi.org/10.1016/j.elecom.2006.06.003}
  {\bibfield  {journal} {\bibinfo  {journal} {Electrochem. Commun.}\ }\textbf
  {\bibinfo {volume} {8}},\ \bibinfo {pages} {1292} (\bibinfo {year}
  {2006})}\BibitemShut {NoStop}%
\bibitem [{\citenamefont {Barpanda}\ \emph {et~al.}(2014)\citenamefont
  {Barpanda}, \citenamefont {Oyama}, \citenamefont {Nishimura}, \citenamefont
  {Chung},\ and\ \citenamefont {Yamada}}]{Barpanda2014}%
  \BibitemOpen
  \bibfield  {author} {\bibinfo {author} {\bibfnamefont {P.}~\bibnamefont
  {Barpanda}}, \bibinfo {author} {\bibfnamefont {G.}~\bibnamefont {Oyama}},
  \bibinfo {author} {\bibfnamefont {S.-i.}\ \bibnamefont {Nishimura}}, \bibinfo
  {author} {\bibfnamefont {S.-C.}\ \bibnamefont {Chung}}, \ and\ \bibinfo
  {author} {\bibfnamefont {A.}~\bibnamefont {Yamada}},\ }\href {\doibase
  https://doi.org/10.1038/ncomms5358} {\bibfield  {journal} {\bibinfo
  {journal} {Nat. Commun.}\ }\textbf {\bibinfo {volume} {5}},\ \bibinfo {pages}
  {4358} (\bibinfo {year} {2014})}\BibitemShut {NoStop}%
\bibitem [{\citenamefont {Melot}\ and\ \citenamefont
  {Tarascon}(2013)}]{Melot2013}%
  \BibitemOpen
  \bibfield  {author} {\bibinfo {author} {\bibfnamefont {B.~C.}\ \bibnamefont
  {Melot}}\ and\ \bibinfo {author} {\bibfnamefont {J.~M.}\ \bibnamefont
  {Tarascon}},\ }\href {\doibase 10.1021/ar300088q} {\bibfield  {journal}
  {\bibinfo  {journal} {Acc. Chem. Res.}\ }\textbf {\bibinfo {volume} {46}},\
  \bibinfo {pages} {1226} (\bibinfo {year} {2013})}\BibitemShut {NoStop}%
\bibitem [{\citenamefont {Kresse}\ and\ \citenamefont
  {Furthmüller}(1996{\natexlab{a}})}]{Kresse1996a}%
  \BibitemOpen
  \bibfield  {author} {\bibinfo {author} {\bibfnamefont {G.}~\bibnamefont
  {Kresse}}\ and\ \bibinfo {author} {\bibfnamefont {J.}~\bibnamefont
  {Furthmüller}},\ }\href
  {https://www.sciencedirect.com/science/article/abs/pii/0927025696000080}
  {\bibfield  {journal} {\bibinfo  {journal} {Comput. Mater. Sci.}\ }\textbf
  {\bibinfo {volume} {6}},\ \bibinfo {pages} {15} (\bibinfo {year}
  {1996}{\natexlab{a}})}\BibitemShut {NoStop}%
\bibitem [{\citenamefont {Kresse}\ and\ \citenamefont
  {Furthmüller}(1996{\natexlab{b}})}]{Kresse1996b}%
  \BibitemOpen
  \bibfield  {author} {\bibinfo {author} {\bibfnamefont {G.}~\bibnamefont
  {Kresse}}\ and\ \bibinfo {author} {\bibfnamefont {J.}~\bibnamefont
  {Furthmüller}},\ }\href
  {https://journals.aps.org/prb/abstract/10.1103/PhysRevB.54.11169} {\bibfield
  {journal} {\bibinfo  {journal} {Phys. Rev. B}\ }\textbf {\bibinfo {volume}
  {54}},\ \bibinfo {pages} {11169} (\bibinfo {year}
  {1996}{\natexlab{b}})}\BibitemShut {NoStop}%
\bibitem [{\citenamefont {Dudarev}\ \emph {et~al.}(1998)\citenamefont
  {Dudarev}, \citenamefont {Botton}, \citenamefont {Savrasov}, \citenamefont
  {Humphreys},\ and\ \citenamefont {Sutton}}]{Dudarev1998}%
  \BibitemOpen
  \bibfield  {author} {\bibinfo {author} {\bibfnamefont {S.}~\bibnamefont
  {Dudarev}}, \bibinfo {author} {\bibfnamefont {G.}~\bibnamefont {Botton}},
  \bibinfo {author} {\bibfnamefont {S.}~\bibnamefont {Savrasov}}, \bibinfo
  {author} {\bibfnamefont {C.}~\bibnamefont {Humphreys}}, \ and\ \bibinfo
  {author} {\bibfnamefont {A.}~\bibnamefont {Sutton}},\ }\href {\doibase
  http://dx.doi.org/10.1039/C7TA10826E} {\bibfield  {journal} {\bibinfo
  {journal} {Phys. Rev. B}\ }\textbf {\bibinfo {volume} {57}},\ \bibinfo
  {pages} {1505} (\bibinfo {year} {1998})}\BibitemShut {NoStop}%
\bibitem [{\citenamefont {Assadi}\ and\ \citenamefont
  {Shigeta}(2018)}]{Assadi2018a}%
  \BibitemOpen
  \bibfield  {author} {\bibinfo {author} {\bibfnamefont {M.~H.~N.}\
  \bibnamefont {Assadi}}\ and\ \bibinfo {author} {\bibfnamefont
  {Y.}~\bibnamefont {Shigeta}},\ }\href {\doibase
  https://doi.org/10.1039/C8RA00576A} {\bibfield  {journal} {\bibinfo
  {journal} {RSC Adv.}\ }\textbf {\bibinfo {volume} {8}},\ \bibinfo {pages}
  {13842} (\bibinfo {year} {2018})}\BibitemShut {NoStop}%
\bibitem [{\citenamefont {Puska}\ \emph {et~al.}(1998)\citenamefont {Puska},
  \citenamefont {Pöykkö}, \citenamefont {Pesola},\ and\ \citenamefont
  {Nieminen}}]{Puska1998}%
  \BibitemOpen
  \bibfield  {author} {\bibinfo {author} {\bibfnamefont {M.~J.}\ \bibnamefont
  {Puska}}, \bibinfo {author} {\bibfnamefont {S.}~\bibnamefont {Pöykkö}},
  \bibinfo {author} {\bibfnamefont {M.}~\bibnamefont {Pesola}}, \ and\ \bibinfo
  {author} {\bibfnamefont {R.~M.}\ \bibnamefont {Nieminen}},\ }\href
  {https://link.aps.org/doi/10.1103/PhysRevB.58.1318} {\bibfield  {journal}
  {\bibinfo  {journal} {Phys. Rev. B}\ }\textbf {\bibinfo {volume} {58}},\
  \bibinfo {pages} {1318} (\bibinfo {year} {1998})}\BibitemShut {NoStop}%
\bibitem [{\citenamefont {Islam}\ and\ \citenamefont
  {Fisher}(2014)}]{Islam2014}%
  \BibitemOpen
  \bibfield  {author} {\bibinfo {author} {\bibfnamefont {M.~S.}\ \bibnamefont
  {Islam}}\ and\ \bibinfo {author} {\bibfnamefont {C.~A.~J.}\ \bibnamefont
  {Fisher}},\ }\href {\doibase 10.1039/C3CS60199D} {\bibfield  {journal}
  {\bibinfo  {journal} {Chem. Soc. Rev.}\ }\textbf {\bibinfo {volume} {43}},\
  \bibinfo {pages} {185} (\bibinfo {year} {2014})}\BibitemShut {NoStop}%
\bibitem [{\citenamefont {Walle}\ and\ \citenamefont
  {Neugebauer}(2004)}]{Walle2004}%
  \BibitemOpen
  \bibfield  {author} {\bibinfo {author} {\bibfnamefont {C.~G. V.~d.}\
  \bibnamefont {Walle}}\ and\ \bibinfo {author} {\bibfnamefont
  {J.}~\bibnamefont {Neugebauer}},\ }\href {\doibase 10.1063/1.1682673}
  {\bibfield  {journal} {\bibinfo  {journal} {J. Appl. Phys.}\ }\textbf
  {\bibinfo {volume} {95}},\ \bibinfo {pages} {3851} (\bibinfo {year}
  {2004})}\BibitemShut {NoStop}%
\bibitem [{\citenamefont {Becke}\ and\ \citenamefont
  {Edgecombe}(1990)}]{Becke1990}%
  \BibitemOpen
  \bibfield  {author} {\bibinfo {author} {\bibfnamefont {A.~D.}\ \bibnamefont
  {Becke}}\ and\ \bibinfo {author} {\bibfnamefont {K.~E.}\ \bibnamefont
  {Edgecombe}},\ }\href {\doibase http://dx.doi.org/10.1063/1.458517}
  {\bibfield  {journal} {\bibinfo  {journal} {J. Chem. Phys.}\ }\textbf
  {\bibinfo {volume} {92}},\ \bibinfo {pages} {5397} (\bibinfo {year}
  {1990})}\BibitemShut {NoStop}%
\bibitem [{\citenamefont {Wu}\ \emph {et~al.}(2015)\citenamefont {Wu},
  \citenamefont {Li}, \citenamefont {Xu}, \citenamefont {Twu}, \citenamefont
  {Liu},\ and\ \citenamefont {Ceder}}]{Wu2015}%
  \BibitemOpen
  \bibfield  {author} {\bibinfo {author} {\bibfnamefont {D.}~\bibnamefont
  {Wu}}, \bibinfo {author} {\bibfnamefont {X.}~\bibnamefont {Li}}, \bibinfo
  {author} {\bibfnamefont {B.}~\bibnamefont {Xu}}, \bibinfo {author}
  {\bibfnamefont {N.}~\bibnamefont {Twu}}, \bibinfo {author} {\bibfnamefont
  {L.}~\bibnamefont {Liu}}, \ and\ \bibinfo {author} {\bibfnamefont
  {G.}~\bibnamefont {Ceder}},\ }\href {\doibase 10.1039/C4EE03045A} {\bibfield
  {journal} {\bibinfo  {journal} {Energy Environ. Sci.}\ }\textbf {\bibinfo
  {volume} {8}},\ \bibinfo {pages} {195} (\bibinfo {year} {2015})}\BibitemShut
  {NoStop}%
\bibitem [{\citenamefont {Bergmann}\ and\ \citenamefont
  {Hinze}(1987)}]{Bergmann1987}%
  \BibitemOpen
  \bibfield  {author} {\bibinfo {author} {\bibfnamefont {D.}~\bibnamefont
  {Bergmann}}\ and\ \bibinfo {author} {\bibfnamefont {J.}~\bibnamefont
  {Hinze}},\ }\href {<Go to ISI>://WOS:A1987K695300007} {\bibfield  {journal}
  {\bibinfo  {journal} {Structure and Bonding}\ }\textbf {\bibinfo {volume}
  {66}},\ \bibinfo {pages} {145} (\bibinfo {year} {1987})}\BibitemShut
  {NoStop}%
\bibitem [{\citenamefont {Politzer}\ and\ \citenamefont
  {Murray}(2002)}]{Politzer2002}%
  \BibitemOpen
  \bibfield  {author} {\bibinfo {author} {\bibfnamefont {P.}~\bibnamefont
  {Politzer}}\ and\ \bibinfo {author} {\bibfnamefont {J.~S.}\ \bibnamefont
  {Murray}},\ }\href {\doibase 10.1007/s00214-002-0363-9} {\bibfield  {journal}
  {\bibinfo  {journal} {Theor. Chem. Acc.}\ }\textbf {\bibinfo {volume}
  {108}},\ \bibinfo {pages} {134} (\bibinfo {year} {2002})}\BibitemShut
  {NoStop}%
\bibitem [{\citenamefont {Jia}\ \emph {et~al.}(2011)\citenamefont {Jia},
  \citenamefont {Zhang}, \citenamefont {Zhang}, \citenamefont {Guo},
  \citenamefont {Zeng},\ and\ \citenamefont {Lin}}]{Jia2011}%
  \BibitemOpen
  \bibfield  {author} {\bibinfo {author} {\bibfnamefont {T.}~\bibnamefont
  {Jia}}, \bibinfo {author} {\bibfnamefont {G.}~\bibnamefont {Zhang}}, \bibinfo
  {author} {\bibfnamefont {X.}~\bibnamefont {Zhang}}, \bibinfo {author}
  {\bibfnamefont {Y.}~\bibnamefont {Guo}}, \bibinfo {author} {\bibfnamefont
  {Z.}~\bibnamefont {Zeng}}, \ and\ \bibinfo {author} {\bibfnamefont {H.~Q.}\
  \bibnamefont {Lin}},\ }\href {\doibase 10.1063/1.3536533} {\bibfield
  {journal} {\bibinfo  {journal} {J. Appl. Phys.}\ }\textbf {\bibinfo {volume}
  {109}},\ \bibinfo {pages} {07E102} (\bibinfo {year} {2011})}\BibitemShut
  {NoStop}%
\bibitem [{\citenamefont {Didier}\ \emph {et~al.}(2011)\citenamefont {Didier},
  \citenamefont {Guignard}, \citenamefont {Denage}, \citenamefont {Szajwaj},
  \citenamefont {Ito}, \citenamefont {Saadoune}, \citenamefont {Darriet},\ and\
  \citenamefont {Delmas}}]{Didier2011}%
  \BibitemOpen
  \bibfield  {author} {\bibinfo {author} {\bibfnamefont {C.}~\bibnamefont
  {Didier}}, \bibinfo {author} {\bibfnamefont {M.}~\bibnamefont {Guignard}},
  \bibinfo {author} {\bibfnamefont {C.}~\bibnamefont {Denage}}, \bibinfo
  {author} {\bibfnamefont {O.}~\bibnamefont {Szajwaj}}, \bibinfo {author}
  {\bibfnamefont {S.}~\bibnamefont {Ito}}, \bibinfo {author} {\bibfnamefont
  {I.}~\bibnamefont {Saadoune}}, \bibinfo {author} {\bibfnamefont
  {J.}~\bibnamefont {Darriet}}, \ and\ \bibinfo {author} {\bibfnamefont
  {C.}~\bibnamefont {Delmas}},\ }\href {\doibase 10.1149/1.3555102} {\bibfield
  {journal} {\bibinfo  {journal} {Electrochem. Solid-State Lett.}\ }\textbf
  {\bibinfo {volume} {14}},\ \bibinfo {pages} {A75} (\bibinfo {year}
  {2011})}\BibitemShut {NoStop}%
\bibitem [{\citenamefont {Yu}\ \emph {et~al.}(2015)\citenamefont {Yu},
  \citenamefont {Park}, \citenamefont {Jung}, \citenamefont {Chung},
  \citenamefont {Aurbach}, \citenamefont {Sun},\ and\ \citenamefont
  {Myung}}]{Yu2015}%
  \BibitemOpen
  \bibfield  {author} {\bibinfo {author} {\bibfnamefont {C.-Y.}\ \bibnamefont
  {Yu}}, \bibinfo {author} {\bibfnamefont {J.-S.}\ \bibnamefont {Park}},
  \bibinfo {author} {\bibfnamefont {H.-G.}\ \bibnamefont {Jung}}, \bibinfo
  {author} {\bibfnamefont {K.-Y.}\ \bibnamefont {Chung}}, \bibinfo {author}
  {\bibfnamefont {D.}~\bibnamefont {Aurbach}}, \bibinfo {author} {\bibfnamefont
  {Y.-K.}\ \bibnamefont {Sun}}, \ and\ \bibinfo {author} {\bibfnamefont
  {S.-T.}\ \bibnamefont {Myung}},\ }\href {\doibase 10.1039/C5EE00695C}
  {\bibfield  {journal} {\bibinfo  {journal} {Energy Environ. Sci.}\ }\textbf
  {\bibinfo {volume} {8}},\ \bibinfo {pages} {2019} (\bibinfo {year}
  {2015})}\BibitemShut {NoStop}%
\bibitem [{\citenamefont {Xia}\ and\ \citenamefont {Dahn}(2011)}]{Xia2011}%
  \BibitemOpen
  \bibfield  {author} {\bibinfo {author} {\bibfnamefont {X.}~\bibnamefont
  {Xia}}\ and\ \bibinfo {author} {\bibfnamefont {J.~R.}\ \bibnamefont {Dahn}},\
  }\href {\doibase 10.1149/2.002201esl} {\bibfield  {journal} {\bibinfo
  {journal} {Electrochem. Solid-State Lett.}\ }\textbf {\bibinfo {volume}
  {15}},\ \bibinfo {pages} {A1} (\bibinfo {year} {2011})}\BibitemShut {NoStop}%
\bibitem [{\citenamefont {Ma}\ \emph {et~al.}(2011)\citenamefont {Ma},
  \citenamefont {Chen},\ and\ \citenamefont {Ceder}}]{Ma2011}%
  \BibitemOpen
  \bibfield  {author} {\bibinfo {author} {\bibfnamefont {X.}~\bibnamefont
  {Ma}}, \bibinfo {author} {\bibfnamefont {H.}~\bibnamefont {Chen}}, \ and\
  \bibinfo {author} {\bibfnamefont {G.}~\bibnamefont {Ceder}},\ }\href
  {\doibase 10.1149/2.035112jes} {\bibfield  {journal} {\bibinfo  {journal} {J.
  Electrochem. Soc.}\ }\textbf {\bibinfo {volume} {158}},\ \bibinfo {pages}
  {A1307} (\bibinfo {year} {2011})}\BibitemShut {NoStop}%
\bibitem [{\citenamefont {Yabuuchi}\ \emph {et~al.}(2012)\citenamefont
  {Yabuuchi}, \citenamefont {Yoshida},\ and\ \citenamefont
  {Komaba}}]{Yabuuchi2012}%
  \BibitemOpen
  \bibfield  {author} {\bibinfo {author} {\bibfnamefont {N.}~\bibnamefont
  {Yabuuchi}}, \bibinfo {author} {\bibfnamefont {H.}~\bibnamefont {Yoshida}}, \
  and\ \bibinfo {author} {\bibfnamefont {S.}~\bibnamefont {Komaba}},\ }\href
  {\doibase 10.5796/electrochemistry.80.716} {\bibfield  {journal} {\bibinfo
  {journal} {Electrochemistry}\ }\textbf {\bibinfo {volume} {80}},\ \bibinfo
  {pages} {716} (\bibinfo {year} {2012})}\BibitemShut {NoStop}%
\bibitem [{\citenamefont {Zhao}\ \emph {et~al.}(2013)\citenamefont {Zhao},
  \citenamefont {Zhao}, \citenamefont {Dimov}, \citenamefont {Okada},\ and\
  \citenamefont {Nishida}}]{Zhao2013}%
  \BibitemOpen
  \bibfield  {author} {\bibinfo {author} {\bibfnamefont {J.}~\bibnamefont
  {Zhao}}, \bibinfo {author} {\bibfnamefont {L.}~\bibnamefont {Zhao}}, \bibinfo
  {author} {\bibfnamefont {N.}~\bibnamefont {Dimov}}, \bibinfo {author}
  {\bibfnamefont {S.}~\bibnamefont {Okada}}, \ and\ \bibinfo {author}
  {\bibfnamefont {T.}~\bibnamefont {Nishida}},\ }\href {\doibase
  10.1149/2.007305jes} {\bibfield  {journal} {\bibinfo  {journal} {J.
  Electrochem. Soc.}\ }\textbf {\bibinfo {volume} {160}},\ \bibinfo {pages}
  {A3077} (\bibinfo {year} {2013})}\BibitemShut {NoStop}%
\bibitem [{\citenamefont {Shibata}\ \emph {et~al.}(2015)\citenamefont
  {Shibata}, \citenamefont {Fukuzumi}, \citenamefont {Kobayashi},\ and\
  \citenamefont {Moritomo}}]{Shibata2015}%
  \BibitemOpen
  \bibfield  {author} {\bibinfo {author} {\bibfnamefont {T.}~\bibnamefont
  {Shibata}}, \bibinfo {author} {\bibfnamefont {Y.}~\bibnamefont {Fukuzumi}},
  \bibinfo {author} {\bibfnamefont {W.}~\bibnamefont {Kobayashi}}, \ and\
  \bibinfo {author} {\bibfnamefont {Y.}~\bibnamefont {Moritomo}},\ }\href
  {\doibase 10.1038/srep09006} {\bibfield  {journal} {\bibinfo  {journal} {Sci.
  Rep.}\ }\textbf {\bibinfo {volume} {5}},\ \bibinfo {pages} {9006} (\bibinfo
  {year} {2015})}\BibitemShut {NoStop}%
\bibitem [{\citenamefont {Vassilaras}\ \emph {et~al.}(2013)\citenamefont
  {Vassilaras}, \citenamefont {Ma}, \citenamefont {Li},\ and\ \citenamefont
  {Ceder}}]{Vassilaras2013}%
  \BibitemOpen
  \bibfield  {author} {\bibinfo {author} {\bibfnamefont {P.}~\bibnamefont
  {Vassilaras}}, \bibinfo {author} {\bibfnamefont {X.}~\bibnamefont {Ma}},
  \bibinfo {author} {\bibfnamefont {X.}~\bibnamefont {Li}}, \ and\ \bibinfo
  {author} {\bibfnamefont {G.}~\bibnamefont {Ceder}},\ }\href {\doibase
  10.1149/2.023302jes} {\bibfield  {journal} {\bibinfo  {journal} {J.
  Electrochem. Soc.}\ }\textbf {\bibinfo {volume} {160}},\ \bibinfo {pages}
  {A207} (\bibinfo {year} {2013})}\BibitemShut {NoStop}%
\bibitem [{\citenamefont {Vitoux}\ \emph {et~al.}(2017)\citenamefont {Vitoux},
  \citenamefont {Guignard}, \citenamefont {Suchomel}, \citenamefont
  {Pramudita}, \citenamefont {Sharma},\ and\ \citenamefont
  {Delmas}}]{Vitoux2017}%
  \BibitemOpen
  \bibfield  {author} {\bibinfo {author} {\bibfnamefont {L.}~\bibnamefont
  {Vitoux}}, \bibinfo {author} {\bibfnamefont {M.}~\bibnamefont {Guignard}},
  \bibinfo {author} {\bibfnamefont {M.~R.}\ \bibnamefont {Suchomel}}, \bibinfo
  {author} {\bibfnamefont {J.~C.}\ \bibnamefont {Pramudita}}, \bibinfo {author}
  {\bibfnamefont {N.}~\bibnamefont {Sharma}}, \ and\ \bibinfo {author}
  {\bibfnamefont {C.}~\bibnamefont {Delmas}},\ }\href {\doibase
  10.1021/acs.chemmater.7b01834} {\bibfield  {journal} {\bibinfo  {journal}
  {Chem. Mater.}\ }\textbf {\bibinfo {volume} {29}},\ \bibinfo {pages} {7243}
  (\bibinfo {year} {2017})}\BibitemShut {NoStop}%
\bibitem [{\citenamefont {Assadi}\ \emph {et~al.}(2018)\citenamefont {Assadi},
  \citenamefont {Okubo}, \citenamefont {Yamada},\ and\ \citenamefont
  {Tateyama}}]{Assadi2018b}%
  \BibitemOpen
  \bibfield  {author} {\bibinfo {author} {\bibfnamefont {M.~H.~N.}\
  \bibnamefont {Assadi}}, \bibinfo {author} {\bibfnamefont {M.}~\bibnamefont
  {Okubo}}, \bibinfo {author} {\bibfnamefont {A.}~\bibnamefont {Yamada}}, \
  and\ \bibinfo {author} {\bibfnamefont {Y.}~\bibnamefont {Tateyama}},\ }\href
  {\doibase http://dx.doi.org/10.1039/C7TA10826E} {\bibfield  {journal}
  {\bibinfo  {journal} {J. Mater. Chem. A}\ }\textbf {\bibinfo {volume} {6}},\
  \bibinfo {pages} {3747} (\bibinfo {year} {2018})}\BibitemShut {NoStop}%
\end{thebibliography}%

\end{document}